\documentclass[12pt,letterpaper]{iopart}

\usepackage{graphicx}       
\usepackage{amsfonts}       
\usepackage{verbatim}       
\usepackage{braket}			
\usepackage{upgreek}		
\usepackage[colorlinks,linkcolor=blue,citecolor=blue,urlcolor=blue]{hyperref}
\usepackage[square, comma, sort&compress, numbers]{natbib}
\usepackage[top=1.25in, bottom=1.25in, left=1in, right=1in]{geometry}
\newcommand{\text}[1]{#1}

\begin{document}

\newcommand{\shorttitle}{
  Spatially uniform gates with near-field microwaves and compensating pulses%
}

\title[\shorttitle]{Spatially uniform single-qubit gate operations with near-field microwaves and composite pulse compensation}

\author{
  Christopher M Shappert$^1$\footnote{Author to whom correspondence should be addressed},
  J True Merrill$^2$,
  K R Brown$^{1}$,
  Jason M Amini$^{1}$,
  Curtis Volin$^1$,
  S Charles Doret$^1$,
  Harley Hayden$^1$,
  C-S Pai$^1$,
  Kenneth R Brown$^{2-4}$, and
  Alexa W Harter$^1$
}

\address{
  $^1$ Georgia Tech Research Institute, Atlanta, GA 30332, USA.
}
\address{
  $^2$ School of Chemistry and Biochemistry, Georgia Institute of Technology, Atlanta, GA 30332, USA
}
\address{
  $^3$ School of Physics, Georgia Institute of Technology, Atlanta, GA 30332, USA
}
\address{
  $^4$ School of Computational Science and Engineering, Georgia Institute of Technology, Atlanta, GA 30332, USA
}

\ead{\href{mailto:chris.shappert@gtri.gatech.edu}{chris.shappert@gtri.gatech.edu}}

\begin{abstract}
We present a microfabricated surface-electrode ion trap with a pair of integrated waveguides that generate a standing microwave field resonant  with the $^{171}$Yb$^+$ hyperfine qubit. The waveguides are engineered to position the wave antinode near the center of the trap, resulting in maximum field amplitude and uniformity along the trap axis. By calibrating the relative amplitudes and phases of the waveguide currents, we can control the polarization of the microwave field to reduce off-resonant coupling to undesired Zeeman sublevels. We demonstrate single-qubit $\pi$-rotations as fast as 1~$\upmu$s with less than 6$\%$ variation in Rabi frequency over an 800~$\upmu$m microwave interaction region. Fully compensating pulse sequences further improve the uniformity of $X$-gates across this interaction region.
\end{abstract}

\pacs{37.10.Ty, 03.67.Lx}

\tableofcontents
\renewcommand{\leftmark}{\shorttitle}

\section{Introduction}
The future development of a fault-tolerant quantum processor will require the storage and manipulation of large numbers of qubits and the ability to execute quantum gates with high fidelity~\cite{DiVincenzo.2000}. Some proposed architectures use trapped atomic ions to store and transport quantum information, and recent years have witnessed significant progress in the control of these systems~\cite{Wineland-LAPL.8.175.2011,Monroe_RMP_2010,Haffner-PhysRep.469.115.2008, Benhelm-NaturePhysics.4.463.2008,Home-Science}. Most trapped-ion experiments utilize laser fields to control  internal and motional quantum states~\cite{Leibfried-RevModPhys.75.281.2003}; however, the fidelity of laser-mediated gates suffers from errors induced by unavoidable spontaneous emission~\cite{Ozeri-PRL.95.030403, Ozeri-PRA.75.042329.2007, Akerman-PRL.109.103601},  pointing instabilities, and variations in the laser frequency, phase, and power~\cite{Wineland-NIST.103.259.1998}. The systems required for laser-mediated gates are also large and complex, making them difficult to scale to the large numbers of qubits required to execute algorithms or simulations of interest.

Alternatively, microwaves can be used to control the ground electronic hyperfine state of the ions, and several experiments have demonstrated high-fidelity single-qubit gates that take advantage of the long natural lifetime of hyperfine qubits and the stability of microwave sources~\cite{Brown-PhysRevA.84.030303.2011,Ospelkaus-Nature.476.181.2011,allcock2013microfabricated,Timoney-Nature.476.185.2011}. The advent of microfabricated ion traps facilitates the use of near-field microwaves, since microwave structures can be integrated directly into the trap. These structures can enable single-ion addressing and multi-qubit gate operations by providing a mechanism to generate field gradients across significantly sub-wavelength inter-ion distances~\cite{Warring-2012,warring2013techniques,Johanning-PRL.102.2009,Ospelkaus-PRL.101.090502.2008,Ospelkaus-Nature.476.181.2011,allcock2013microfabricated,Wang-APL.94.094103.2009}. In addition, ions trapped in the near-field can experience a large field amplitude resulting in fast gate operations.

Here we present a microfabricated surface-electrode ion trap with integrated microwave waveguides fabricated with standard very large-scale integration (VLSI) techniques. The waveguides are designed to place an antinode of a 12.64~GHz standing microwave field, suitable for coupling the ground state hyperfine levels of $^{171}$Yb$^{+}$, at the center of the ion trap. The microwaves are polarization tunable, so that we can optimize coupling to a desired Zeeman transition while suppressing off-resonant coupling to neighboring transitions. With the ion located 59~$\upmu$m above the trap surface, a 0.037~mT field amplitude allows execution of sub-microsecond $\pi$-rotations. The intrinsic position dependence of the standing microwave field causes the Rabi frequency to vary by less than 6$\%$ over an 800~$\upmu$m linear span. We correct for this residual non-uniformity by executing gates with fully compensating broadband pulse sequences.

\begin{figure}
  \begin{centering}
    \includegraphics{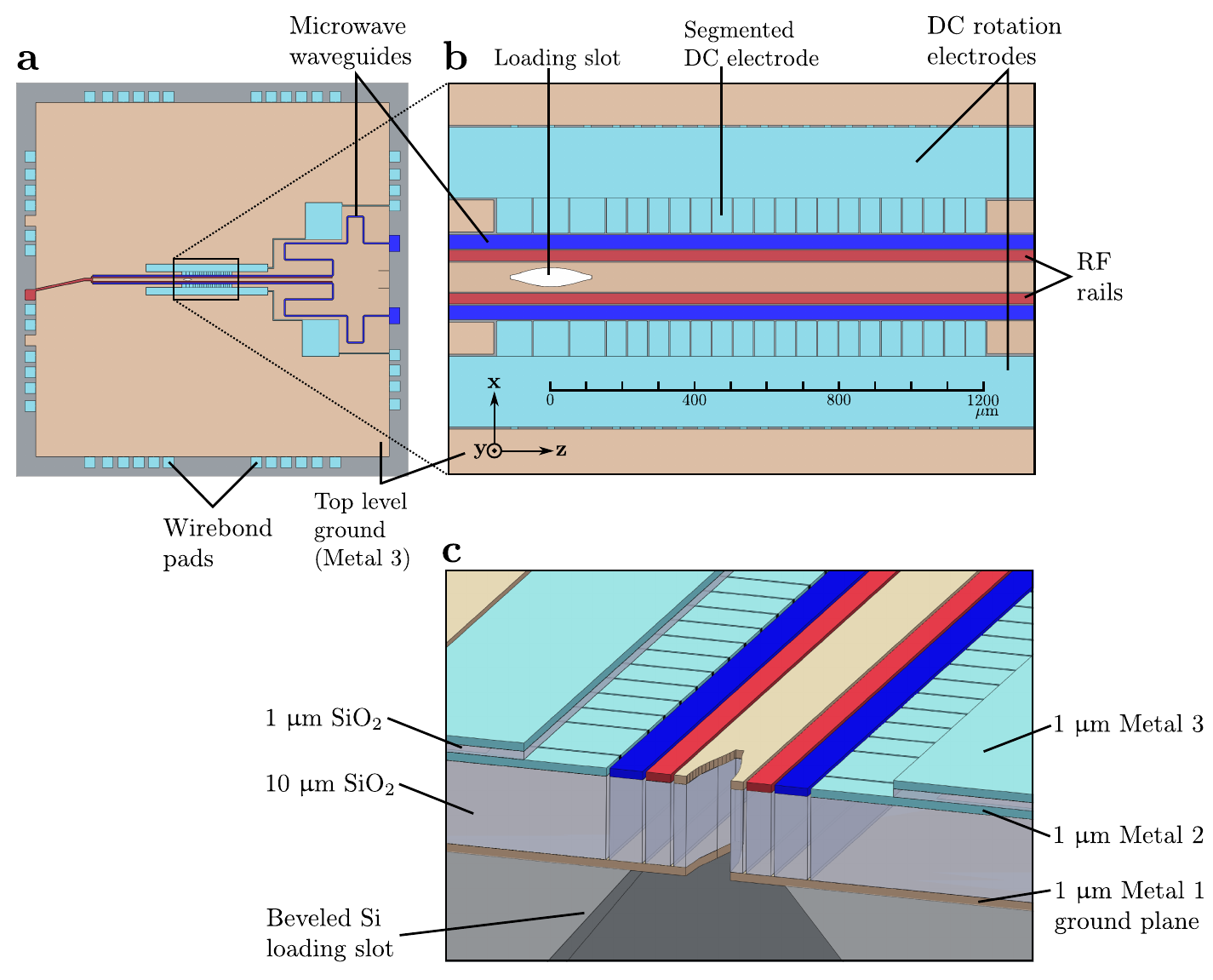}
    \caption{
     (a) A schematic of the $11\times11$~mm$^2$ silicon chip (b) The active trapping region, showing RF and DC trapping electrodes, the shaped loading slot, and the on-chip microwave waveguides. (c) A cross section showing internal layers (the vertical direction is scaled by $10\times$ for visual clarity).
      \label{fig:1}
    }
  \end{centering}
\end{figure}

\section{Trap design}
\label{sec:design}

\subsection{Trapping structures}

The trap conforms to a symmetric five-wire surface-electrode Paul trap geometry~\cite{Chiaverini-QIC.96.253003.2005} fabricated on a $11\times11$~mm$^2$ silicon die (figure~\ref{fig:1}a) similar to the designs reported in~\cite{Doret-NJP.14.073012.2012, Merrill-NJP.13.103005.2011, wright2013reliable}.  Electrodes etched into three sputtered aluminum layers are separated by insulating silicon dioxide films.  Radio-frequency (RF) potentials applied to two parallel electrodes provide radial ion confinement (in the $x$-$y$ plane, figure~\ref{fig:1}b) 59~$\upmu$m above the Metal 2 layer (figure~\ref{fig:1}c). Quasi-static potentials applied to segmented DC electrodes confine and transport ions along $z$. The RF electrodes are 30~$\upmu$m wide along $x$ with an inner edge-to-edge separation of 92~$\upmu$m. The segmented DC electrodes are 56~$\upmu$m wide along $z$ except for six 100~$\upmu$m wide electrodes bordering the loading slot. Two additional DC electrodes traversing the entire length of the trapping region are used to apply uniform $x$-$y$ fields and to rotate the radial principal axes. Each electrode is separated from neighboring conductors by 4~$\upmu$m gaps. Each DC electrode incorporates a 60~pF plate capacitor (1~mm$^2$ area) to filter unwanted RF pickup~\cite{Doret-NJP.14.073012.2012}. A loading slot allows a thermal beam of neutral Yb to reach the trapping volume from an oven located below the trap.

\subsection{Integrated waveguides}
The trap includes a pair of conductor-backed coplanar waveguides that generate local microwave magnetic fields. Each waveguide includes a 40~$\upmu$m wide electrode with 4~$\upmu$m gaps to neighboring conductors, and 10~$\upmu$m of SiO$_2$ separate the coplanar layer from the ground plane below (figure~\ref{fig:1}b). The waveguides support a $\omega_\mathrm{mw} = 2\pi\times12.64$~GHz quasi-TEM guided mode resonant with the hyperfine splitting between the $F = 0$ and $F = 1$ manifolds in the ${}^2\mbox{S}_{1/2}$ ground state of $^{171}$Yb$^+$ (see figure~\ref{fig:3}a). In the ideal case, currents in each waveguide generate a magnetic field along the trapping axis
\begin{eqnarray}
	 \vec{B}_k(z,t)=\left ( \hat{x}\beta_{x,k} + \hat{y}\beta_{y,k}\right )I_{k}(z)\cos{( \omega_\mathrm{mw}t + \phi_{k})}
	 \label{eq:bfield}
\end{eqnarray}
where $k=1,2$ is an index for the two waveguides, $\phi_{k}$ is the phase of the microwave current source, $I_{k}(z)$ is the current in each waveguide, and $\beta_{x,1}=\beta_{x,2}\simeq 0.08$~mT/A and $\beta_{y,1}=-\beta_{y,2} \simeq 0.17$~mT/A are properties of the waveguide mode. $\beta_{y,1}=-\beta_{y,2}$ due to the symmetric placement of the waveguides around the trapping axis. The waveguides terminate in an open circuit at a position that is approximately a quarter-wavelength from the trap center, which produces a standing wave field with maximum amplitude and uniformity in the gate region. A smaller traveling wave component also exists due to on-chip attenuation that generates an amplitude difference between forward and backward propagating waves. 

\begin{figure}
  \begin{centering}
    \includegraphics{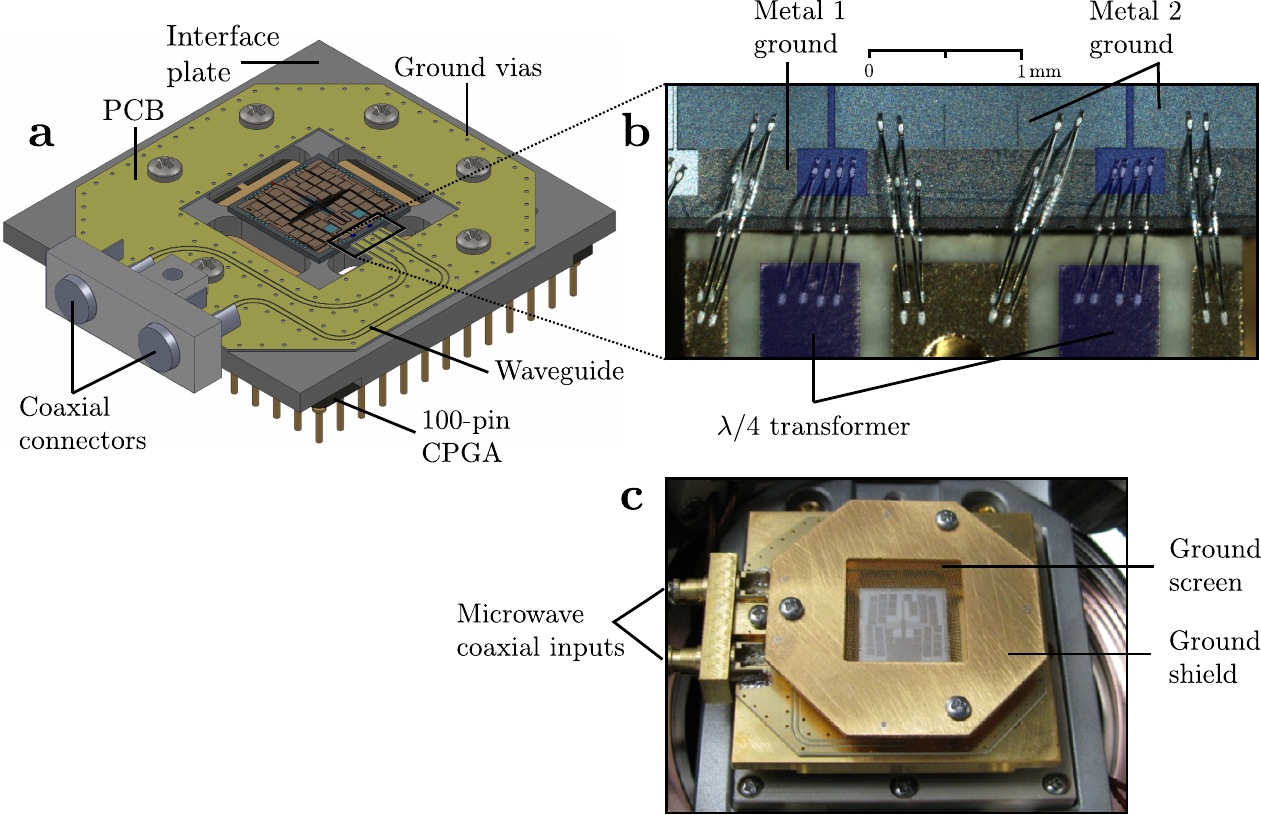}
    \caption{
      Components for trap packaging and microwave power delivery.  (a) Schematic of a packaged trap with top ground screen removed for clarity. (b) Colorized photograph of the wirebond interface between the PCB and the trap die. (c) Photo of the installed device with coaxial microwave connections.
      \label{fig:2}
    }
  \end{centering}
\end{figure}

Far from the trap center, the waveguides meander to fit a complete wavelength on the chip and then terminate on wirebond pads at the edge of the chip. Extending the waveguides to a full wavelength in this way places a current node at the wirebond pads and reduces the potential for resistive power loss in the connections. A series of 25.4~$\upmu$m diameter aluminum wirebonds connect the chip waveguides to two PCB waveguides that route microwaves from the edge of the trap package (figure~\ref{fig:2}). Connections between the PCB top level ground and the on-chip Metal 1 and Metal 2 ground planes are symmetric about each microwave electrode. Quarter-wave transformers match the 50~ohm impedance of the PCB waveguide to the 27~ohm on-chip characteristic impedance. The PCB is fabricated using a 254~$\upmu$m thick Rogers 4350B substrate with two 18~$\upmu$m thick copper foil conductive layers and a 3-6~$\upmu$m electroless nickel immersion gold (ENIG) finish. The skin depth in the PCB at $\omega_\mathrm{mw}$ is comparable to the thickness of the lossy nickel layer resulting in $\approx 3$~dB of power loss between the microwave connector and wirebonds. Much of this power loss could be recovered by replacing the ENIG finish with an electroplated gold surface.

\section{Trap performance}
\label{sec:performance}

\subsection{%
  \texorpdfstring{Microwave control of $^{171}$Yb$^+$ hyperfine qubits}%
  {Microwave control of 171Yb+ hyperfine qubits}%
}
\label{sec:control}
\begin{figure}
  \begin{centering}
   \includegraphics{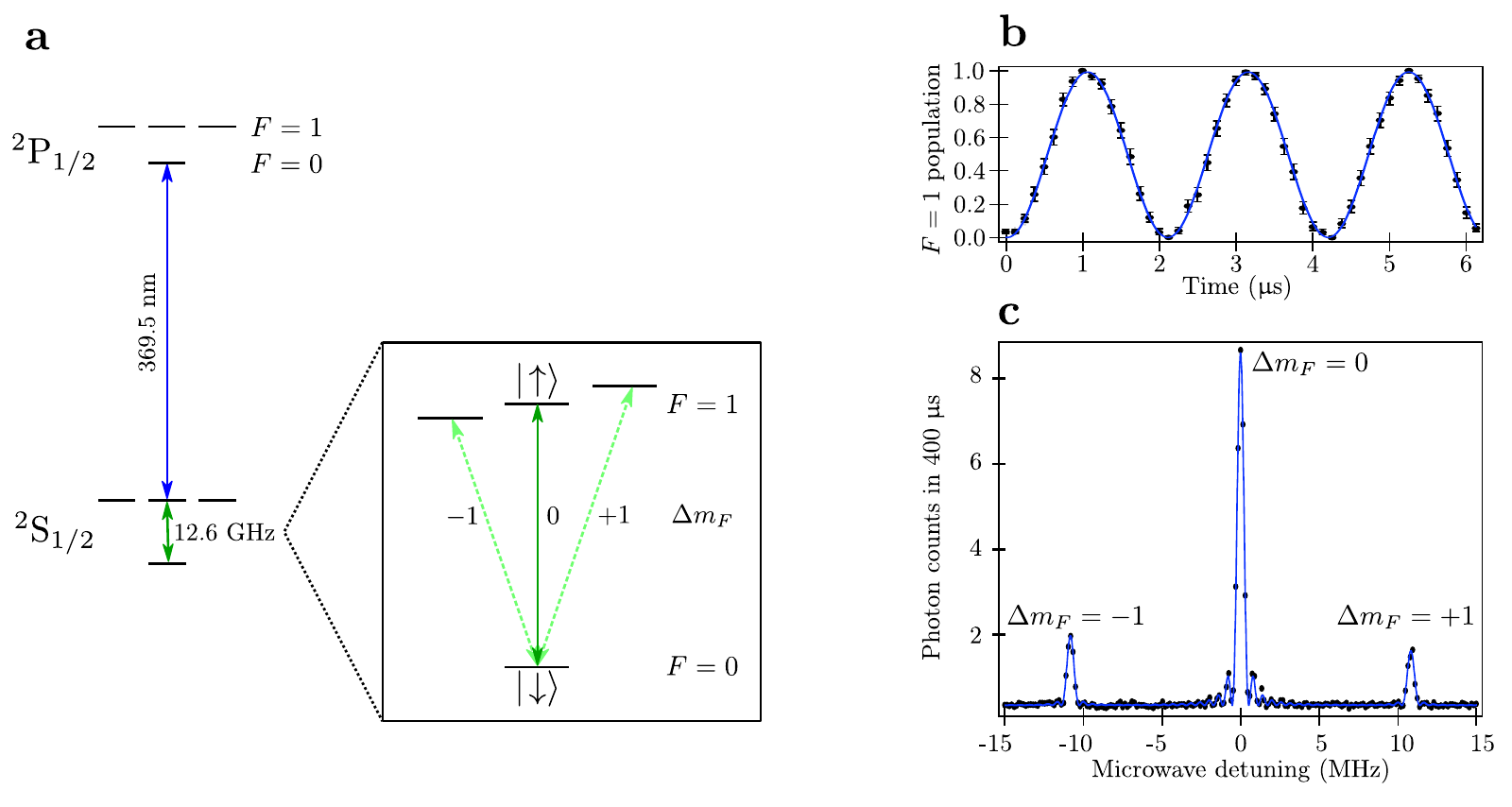}
    \caption{
      (a) Simplified energy level structure of the $^{171} \mbox{Yb}^{+}$ ion (not to scale).  The qubit uses the $|F=1, m_F=0\rangle \equiv  \ket{\uparrow}$ and $|F=0, m_F=0\rangle \equiv \ket{\downarrow}$ clock states.  During experiments, a 12.64~GHz microwave field produced by on-chip waveguides couples hyperfine states in the $^2\mbox{S}_{1/2}$ manifold.  The ${}^2\mbox{S}_{1/2} \rightarrow {}^2\mbox{P}_{1/2}$ transition at 369.5~nm is used for Doppler-cooling, qubit initialization, and qubit detection~\cite{Olmschenk-PRA.76.052314.2007}.  (b) Rabi oscillations between the qubit states induced by application of microwaves ($z=300~\upmu$m). (c) Resolved hyperfine transitions between $^2\mbox{S}_{1/2}$ sublevels. 
      \label{fig:3}
    }
  \end{centering}
\end{figure}

Figure~\ref{fig:3}a shows the transitions in the ground state hyperfine manifold of $^{171}$Yb$^+$ addressed by the microwave field.  A static 0.74~mT field along $y$ defines the quantization axis and lifts the degeneracy of the $F=1$ triplet. We select the clock states $|^2\mbox{S}_{1/2}, F = 1, m_F = 0\rangle \equiv \ket{\uparrow}$ and $|^2\mbox{S}_{1/2}, F = 0,m_F = 0 \rangle \equiv \ket{\downarrow}$ as the qubit states. We observe Rabi oscillations at a frequency $\Omega = 2 \pi \times 0.49$~MHz (figure~\ref{fig:3}b) by optically pumping into $\ket{\downarrow}$, applying microwave power for a variable interval of time, and measuring the resulting population transfer into the $F = 1$ manifold through state-selective fluorescence of the 369.5~nm cycling transition (figure~\ref{fig:3}a)~\cite{Olmschenk-PRA.76.052314.2007}. 

Currents in each waveguide generate an oscillating magnetic field which contains both a ${\pi}-$polarized component that couples to the $\Delta m_F = 0$ clock transition and also a transverse-polarized component that couples to the $\Delta m_F = \pm 1$ transitions. Figure~\ref{fig:3}c shows the hyperfine spectrum measured with microwave power applied to a single waveguide. From the relative Rabi frequencies of the transitions, we estimate the ratio of polarization components for each waveguide as $|\beta_{x,k}| / |\beta_{y,k}| \approx 0.46$.

\begin{figure}
  \begin{centering}
    \includegraphics{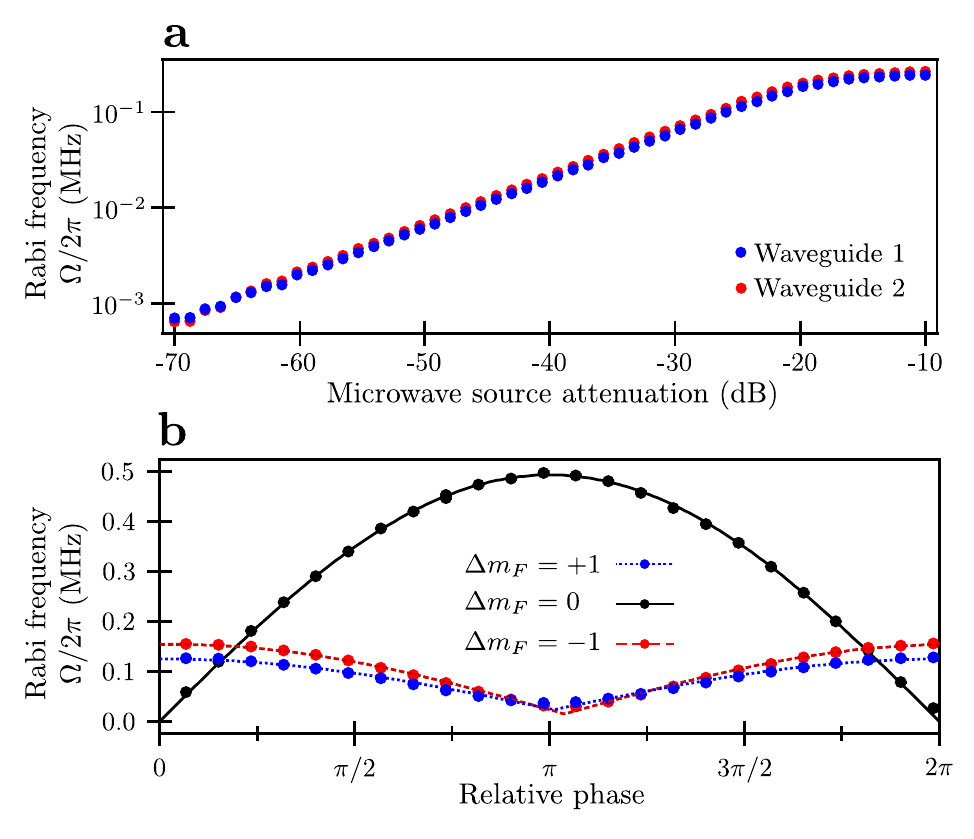}
    \caption{(a) Plot of the qubit Rabi frequency versus microwave source attenuation, obtained with a resonant field produced by a single waveguide.  The nonlinearity at high source power is caused by amplifier saturation. (b) Polarization control via microwave power balancing and phase tuning between the two waveguides.  The solid curves are fits to the magnitude of a sinusoid. 
       \label{fig:4}
     }
   \end{centering}
\end{figure}

The polarization of the near-field microwaves may be controlled by adjusting the relative amplitude and phase of the microwave currents in the two waveguides. In particular, the polarization may be aligned along the quantization axis, thereby maximizing the qubit transition Rabi frequency while  also suppressing off-resonant $\Delta m_F = \pm 1$ transitions. The active and passive microwave components supplying energy to these electrodes are not perfectly power-balanced and phase-matched. We calibrate the microwave sources by first driving each waveguide independently to map the relationship between source power and Rabi frequency (figure~\ref{fig:4}a).  Once the field amplitudes from the waveguides have been equalized, the relative phase between microwave currents $\phi_r$ can be varied to produce an arbitrary linear polarization in the $x$--$y$ plane. Figure~\ref{fig:4}b shows the resonant Rabi frequency for each of the three transitions as $\phi_r$ is varied, demonstrating the desired suppression of the $\Delta m_F = \pm 1$ transitions at $\phi_r=\pi$. We suspect that the mismatch between the $\Delta m_F = \pm 1$ curves in figure~\ref{fig:4}b is caused by a frequency dependence of the microwave system output power.  

\subsection{Microwave field uniformity and compensation}
\label{sec:compensation}

\begin{figure}
   \begin{centering}
    \includegraphics{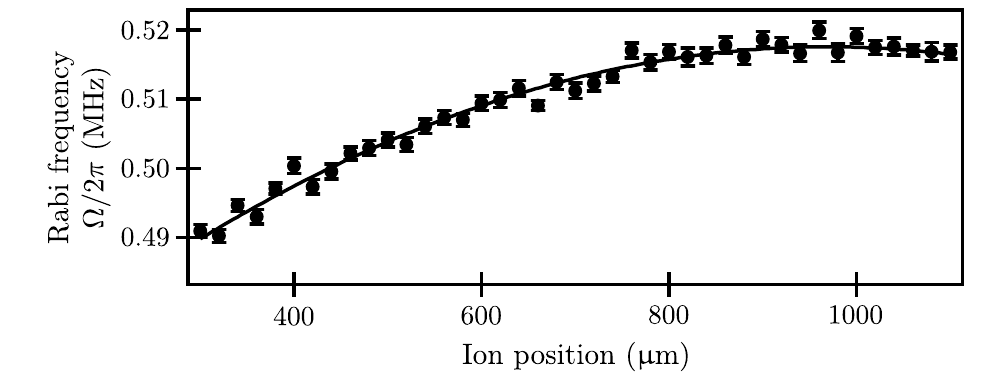}
     \caption{Measured Rabi frequency of the qubit transition as a function of the axial distance from the load slot center. The solid line is a quadratic fit to the data which indicates a maximum microwave field at $z=957$~$\upmu$m.
       \label{fig:5}
     }
   \end{centering}
 \end{figure}

The standing wave current in the waveguides produces a microwave field with non-uniform amplitude along the trap axis.  Figure~\ref{fig:5} shows the measured Rabi frequency of the qubit transition at several positions along the trap axis.  We observe a maximum Rabi frequency of $2\pi \times 0.52$~MHz, corresponding to a field amplitude of 0.037~mT, located $z_0 = 957$~$\upmu$m from the loading slot center. Finite element calculations predict an antinode location at $z = 895$~$\upmu$m in reasonable agreement with the experiment. These models indicate that the maximum field corresponds to a local current in each electrode of $I_z(z_0)\approx0.1$~A.

The microwave field non-uniformity appears as an effective amplitude error when implementing global single-qubit rotations on multiple ions located at different positions in the trap. To improve single-qubit gate uniformity, we implement global gates using broadband compensating pulse sequences~\cite{Levitt1986, Merrill-2012}. 
This technique replaces a simple pulse with a sequence of pulses whose phases are chosen to nearly cancel the effective microwave amplitude error. The excitation profiles of such pulse sequences enable global rotations on many qubits, although the microwave amplitude may differ significantly between distant ions.

\begin{figure}
   \begin{centering}
    \includegraphics{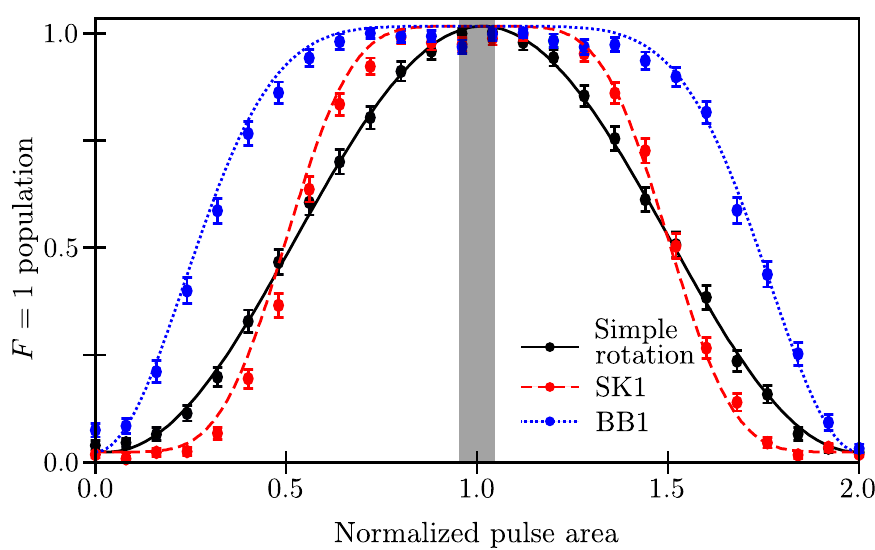}
     \caption{Population transfer into the $F = 1$ manifold after applying a logical $X$ produced by simple pulses (black), SK1 (red), and BB1 (blue) sequences.  The pulse areas are uniformly scaled by adjusting the pulse lengths.  The curves are the signals predicted by theory, adjusted to account for a known qubit-detection error.  The shaded area encloses the range of systematic microwave amplitude error observed over the entire trapping region.
       \label{fig:6}
     }
   \end{centering}
 \end{figure}

Here we demonstrate the error-canceling properties of composite $X$-gates constructed from the first-order SK1~\cite{Brown-PRA.70.052318.2004, Brown-PRA.72.039905.2005} and second-order BB1~\cite{Wimperis-JMR.109.221.1994} sequences.  The experiments prepare the qubit in $\ket{\downarrow}$, apply a logical $X$-gate, and then measure the population in the $F = 1$ manifold.  To simulate the effect of systematic over/under rotations, we uniformly scale the pulse areas of every pulse in the sequence by adjusting the pulse duration.  Figure~\ref{fig:6} plots the measured excitation profiles produced by compensated gates, overlayed on the signal predicted by theory.  
Assuming the field non-uniformity is the sole source of error, we calculate theoretical fidelities of $X$-gates. For the 6\% amplitude deviation observed at $z=300~\upmu$m, a simple rotation performs a global $X$-gate with a minimum fidelity $\mathcal{F} \geq 0.995$, whereas SK1 and BB1 perform the same gate with minimum fidelities of $\mathcal{F} \geq 1 - 1.5 \times 10^{-4}$ and $\mathcal{F} \geq 1 - 2.2 \times 10^{-7}$ respectively. 

 \begin{figure}
   \begin{centering}
    \includegraphics{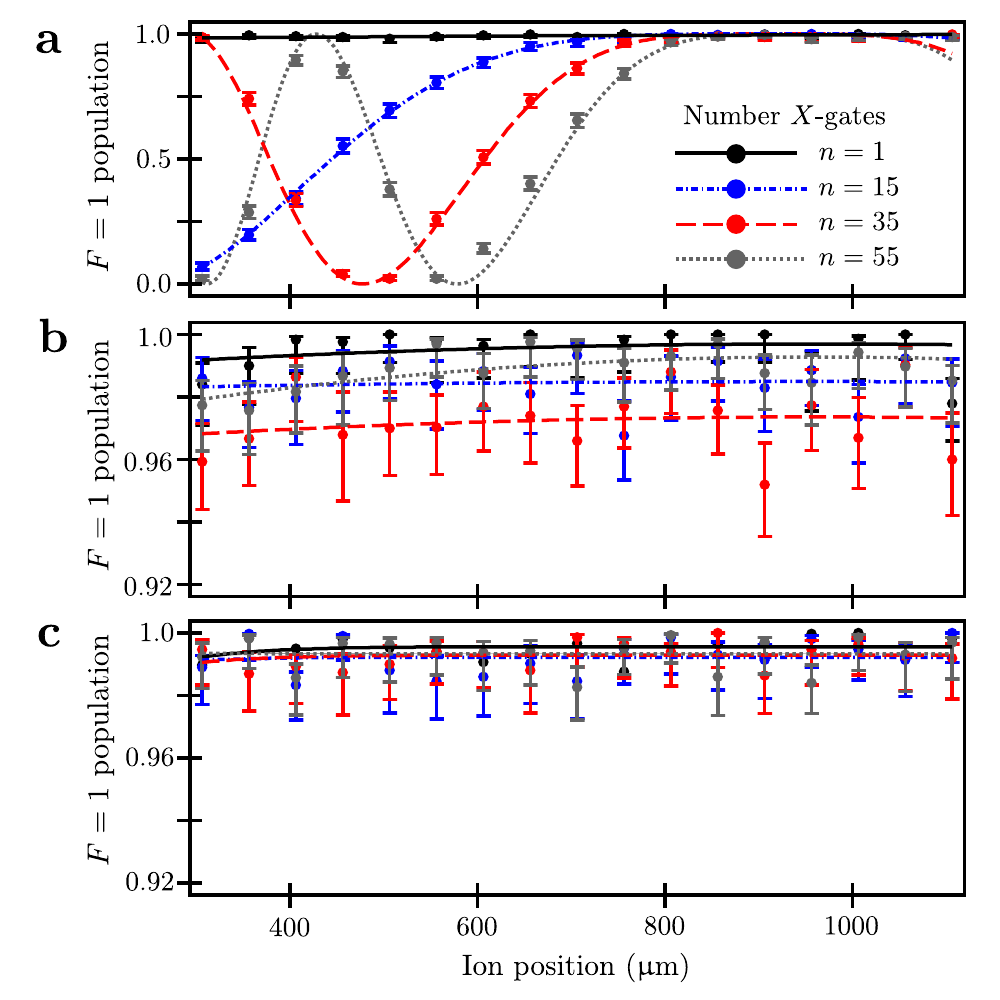}
     \caption{Excited state population fraction after application of $n$ sequential Pauli $X$-gates, with each gate implemented via a (a) simple rotation operator with no compensating pulse sequence, (b) SK1 composite pulse sequence, or (c)  BB1 composite pulse sequence.  Lines are fits to curves predicted by theory given the position-dependent Rabi frequency shown in figure~\ref{fig:5}.
       \label{fig:7}
     }
   \end{centering}
 \end{figure}

As a demonstration of uniform global gates, we perform an experiment where $n$ sequential logical $X$-gates are applied to a qubit initialized in $\ket{\downarrow}$.  We calibrate gate times so that an ion located at the microwave amplitude maximum ($z_0 = 957$~$\upmu$m) experiences nearly perfect rotations.  Qubits displaced from the field maximum rotate at lower Rabi frequencies, acquiring an under-rotation error that accumulates as $n$ increases.  We measure the $F = 1$ population as a function of ion axial position and number of sequential $X$-gates.  For $X$-gates implemented by simple rotations we observe fringes (figure~\ref{fig:7}a) arising from the local qubit falling behind by an entire Rabi cycle relative to the maximal Rabi frequency.  Instead, when implementing SK1 or BB1 pulses (figures~\ref{fig:7}b and \ref{fig:7}c) the error accumulates so slowly that the excitation profile remains flat over the trapping region after $n = 55$ logical $X$-gates.  Our ability to resolve fringe structure in these cases is currently limited by systematic state-preparation and measurement errors and by the number of simple pulse operations we can implement.

We analytically calculate the fidelity scaling of the sequential logical $X$-gates as a function of the microwave field strength.  For simple rotations, the fidelity drops as $\mathcal{F} = | \cos[ \epsilon(z) \pi n / 2 ] |$, where $\epsilon(z) = [ \Omega(z) - \Omega(z_0) ] / \Omega(z_0)$ is the fractional difference in Rabi frequencies between the ion location $z$ and the field maximum.  
For SK1 pulses the fidelity scales as $\mathcal{F} = |1 - \frac{15}{128} \pi^4 \epsilon(z)^4  n + O( \epsilon(z)^6)|$, and for BB1 the fidelity scales as $\mathcal{F} = |1 - \frac{5}{1024} \pi^6 \epsilon(z)^6  n + O( \epsilon(z)^8)|$.


\section{Conclusion}

We developed a microfabricated surface-electrode ion trap with integrated microwave waveguides that performs arbitrary single-qubit gates on the $^{171}$Yb$^{+}$ hyperfine qubit. The polarization of the local microwave field can be tuned to minimize off-resonant coupling to adjacent Zeeman sub-levels. This suppression of off-resonant coupling will be most useful when Rabi frequencies are large compared to the Zeeman splittings, thus causing reduced spectral resolution of neighboring transitions. We demonstrate the use of fully compensating pulse sequences to reduce single-qubit gate variation caused by a spatially non-uniform microwave amplitude. Similar passband and narrowband pulse sequences~\cite{Merrill-2012} could be used to exploit microwave amplitude variations to enable single-ion addressing without requiring perfect field suppression at the location of neighboring ions.


\ack
\addcontentsline{toc}{section}{Acknowledgments}
This material is based upon work supported by the Office of the Director of National
Intelligence (ODNI), Intelligence Advanced Research Projects Activity (IARPA) under U.S. Army Research Office (ARO) contract
W911NF1010231.
All statements of fact, opinion, or conclusions contained herein are those of the authors and should not be construed as representing the official views or policies of IARPA, the ODNI, or the U.S. Government. 


\section*{Appendix: Methods}
\label{sec:methods}
\addcontentsline{toc}{section}{Appendix: Methods}

\subsection*{Microwave delivery system}
\begin{figure}
  \begin{centering}
   \includegraphics{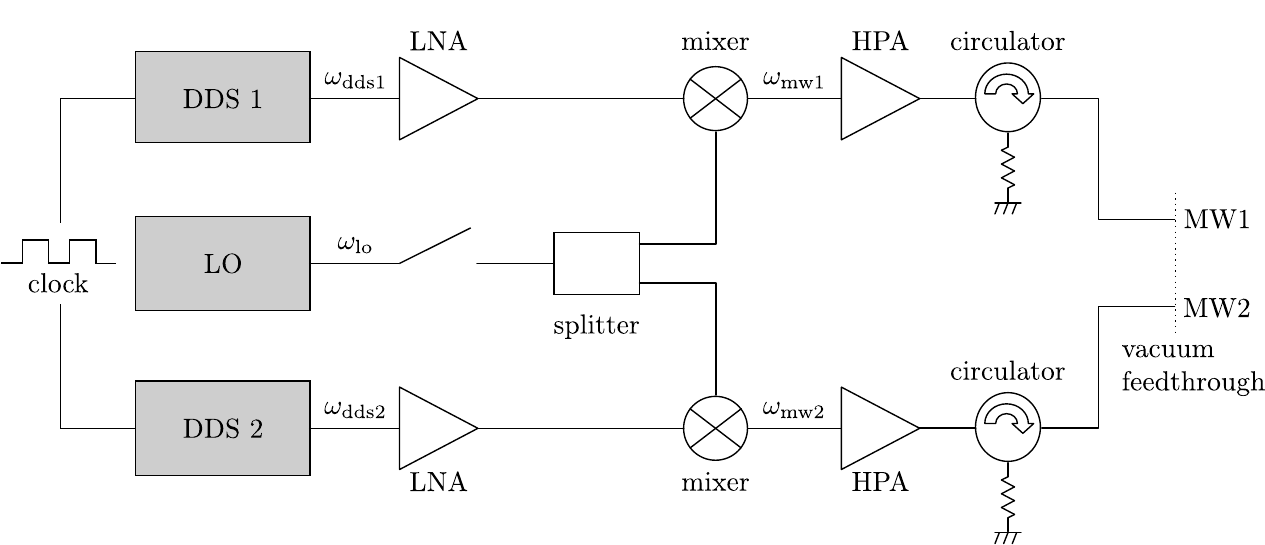}
    \caption{
      Schematic of microwave delivery system connected to the outside of the UHV chamber. The two DDS sources are mixed with a stable local oscillator and provide independent control over the amplitude, phase, and frequency ($\omega_\mathrm{mw}=\omega_\mathrm{lo}\pm \omega_\mathrm{dds}$) of the microwave signal applied to the trap waveguides, though during typical operation $\omega_\mathrm{dds1}=\omega_\mathrm{dds2}$. DDS = direct digital synthesizer, LO = local oscillator, LNA = low noise amplifier, HPA = high power amplifier, MW=microwave.
      \label{fig:9}
    }
  \end{centering}
\end{figure}
A schematic of the microwave supply system is shown in figure~\ref{fig:9}. An Agilent 83623B Swept-Signal Generator provides a stable microwave local oscillator (LO) signal that is approximately 300~MHz detuned from the qubit resonant frequency near 12.64~GHz. Delivery of this signal is controlled by an American Microwave SW-218 high-speed RF switch.
The signal is split to supply the local oscillator ports of two frequency mixers.
Two separate direct digital synthesis (DDS) boards with synchronized clocks supply the intermediate frequency (IF) signals
near 300~MHz.
The DDS outputs have independently controllable amplitude, phase, and frequency and these signals are amplified prior to mixing with the LO signal.
The mixers reject the carrier frequency and produce two RF sidebands at $\omega_\mathrm{mw}=\omega_\mathrm{lo}\pm \omega_\mathrm{dds}$ where one of these sidebands is tuned to the frequency of the desired hyperfine transition, while the other is far off resonance. Both signals are amplified in separate Mini-Circuits ZVE-3W-183+ amplifiers and routed to a coaxial feedthrough port on the UHV chamber.
Inside the chamber the microwaves are carried by two coaxial lines with Kapton dielectric and braided conductor, selected for UHV compatibility. The combined power loss in the feedthrough and Kapton cables is $\approx 5.3$~dB.

\subsection*{Trap packaging}
We mount chips to a gold electroplated stainless steel plate that also serves as a structurally rigid platform for mounting other components, including microwave connectors, printed circuit board (PCB), grounded screen, and a 100-pin ceramic pin-grid array (CPGA).  The chip is bonded to the interface plate with electrically conductive epoxy (Epoxy Technology H21D) which accommodates the differential thermal expansion between the stainless steel mounting plate and the silicon chip during vacuum bake-out.  Wire bonds at the chip perimeter establish electrical connections to the CPGA to provide both DC and RF trapping potentials, while wire bonds to the PCB supply microwave energy to the chip (figure~\ref{fig:2}b).  To minimize scattered laser light, wire bonds are excluded from regions where cross-chip laser access is required.






\end{document}